\documentclass[sigconf, noacm]{acmart}
\usepackage{newtxtext} 
\usepackage{newtxmath} 

\usepackage{graphicx}
\usepackage{cleveref}
\usepackage{comment}
\title{A GPU-Accelerated RAG-Based Telegram Assistant for Supporting Parallel Processing Students}
\author{Guy Tel-Zur}
\affiliation{%
	\institution{Ben-Gurion University of the Negev}
	\city{Beer Sheva}
	\country{Israel}
}

\setcopyright{none}
\renewcommand\footnotetextcopyrightpermission[1]{}
\acmConference[]{}{}{}

\begin{document}

	\begin{abstract}
		This project addresses a critical pedagogical need: offering students continuous, on-demand academic assistance beyond conventional reception hours. I present a domain-specific Retrieval-Augmented Generation (RAG) system powered by a quantized Mistral-7B Instruct model\cite{website:mistral} and deployed as a Telegram bot\cite{website:telegram}. The assistant enhances learning by delivering real-time, personalized responses aligned with the "Introduction to Parallel Processing" course materials [1]. GPU acceleration significantly improves inference latency, enabling practical deployment on consumer hardware.
		This approach demonstrates how consumer GPUs can enable affordable, private, and effective AI tutoring for HPC education.
	\end{abstract}
	
	\maketitle
	\section{Introduction}
	Large Language Models (LLMs) have revolutionized human- computer interaction. However, deploying these systems in a privacy- preserving and cost-effective way remains a challenge. 
	This paper describes the development of a local Retrieval-Augmented Generation (RAG) assistant using a quantized version of Mistral-7B model. The system is deployed as a Telegram bot to support students enrolled in the "Introduction to Parallel Processing" course\cite{website:ipp}. It runs entirely on a local machine with a consumer GPU, ensuring both privacy and responsiveness.
	The term RAG was coined by Lewis, P. et al\cite{Lewis2020}. In the words of Vinton Cerf:
	"RAG systems connect LLMs to external, verifiable knowledge bases, allowing them to ground their responses 	in current, curated information rather
	than relying solely on their training data. This dramatically reduces the
	production of factual errors and hallucinations. Some research indicates
	improvements of $42\%-68\%$ and even higher in specific domains, such as
	medical, AI when paired with trusted sources"\cite{Cerf2025}.
	
	\section{Project description}
	I have been teaching Parallel Processing for more than 20 years. In 2014 I described the course called "An Introduction to Parallel Processing" in the EduHPC 2014 workshop\cite{Tel-Zur2014}. The current course web site is available at \cite{website:ipp}. Every lecturer is committed conduct reception hours, on a weekly basis, for answering students questions. sometimes this time slot isn't enough  especially toward the final examinations dates where the students are more focused on learning toward the exams and have more questions to ponder. In addition, some of the student may feel shy and will avoid asking questions during these hours. Today when AI is becoming so advanced it is possible to provide a solution to these needs where a smart agent can be available continuously  $24\times7\times365$. 
	When this project idea triggered my imagination, a few months ago, developing a smart agent was still a complicated task. However, with the accelerating pace of AI it is now becoming a very popular topic and there are many alternative ways to implement smart agents. However, the project that I describe here still has a few unique features:
	\newline $\bullet$ This project is built using only open-source tool. This means that there are no licensing issues, no payments, and everything is running on a stand alone computer so that privacy is kept if that is important to the users.
	\newline $\bullet$ This smart agent uses a Telegram interface which is available from any platform (desktop, mobile phone, tablet, etc').
	\newline $\bullet$ The smart bot can run on a commodity computer preferably with a Graphics Processing Unit (GPU). In this project I use an ASUS TUF F17 laptop with 32GB RAM and an Nvidia GeForce RTX 4060 GPU, and the response time was found to be reasonable.
	In addition, the smart agent project simplicity allows it to be implemented as an educational assignment in AI related courses in addition to the main goal of helping students.
	
	\section{Deployment}
    The course slides were merged into a single PDF file and together with the electronic version of the course textbook\cite{wilkinson} served as the knowledge base for the smart-agent.
    A document preparation pipeline is next in order to
    build a searchable knowledge base for the RAG system.
    
\begin{figure*}[p]
    	\centering
    	\includegraphics[width=1.3\textwidth, angle=90]{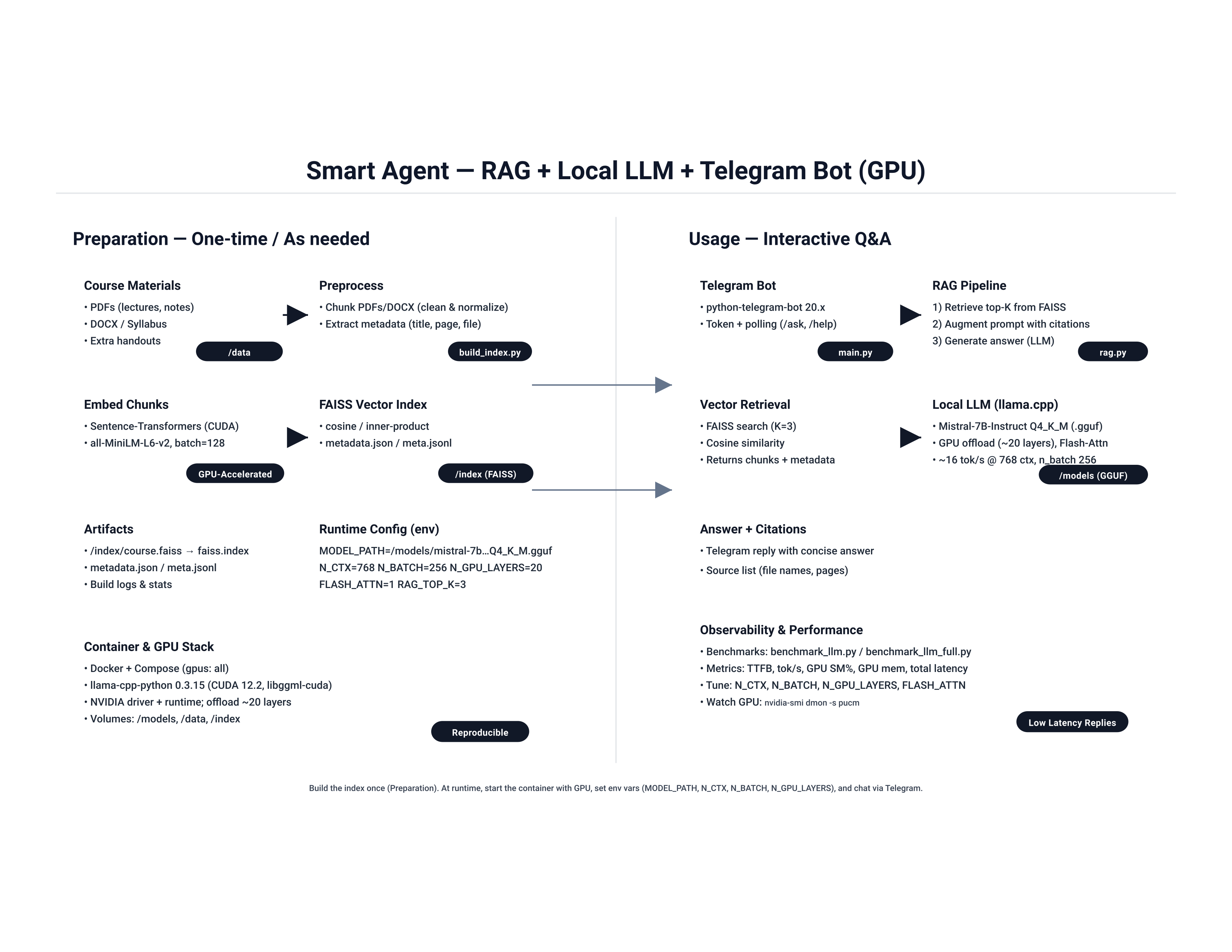}
    	\caption{The smart agent project block diagram.}
    	\label{fig:smart-agent-architecture}
    \end{figure*}

\begin{figure}[htbp]
    	\centering
    	\includegraphics[scale=0.20]{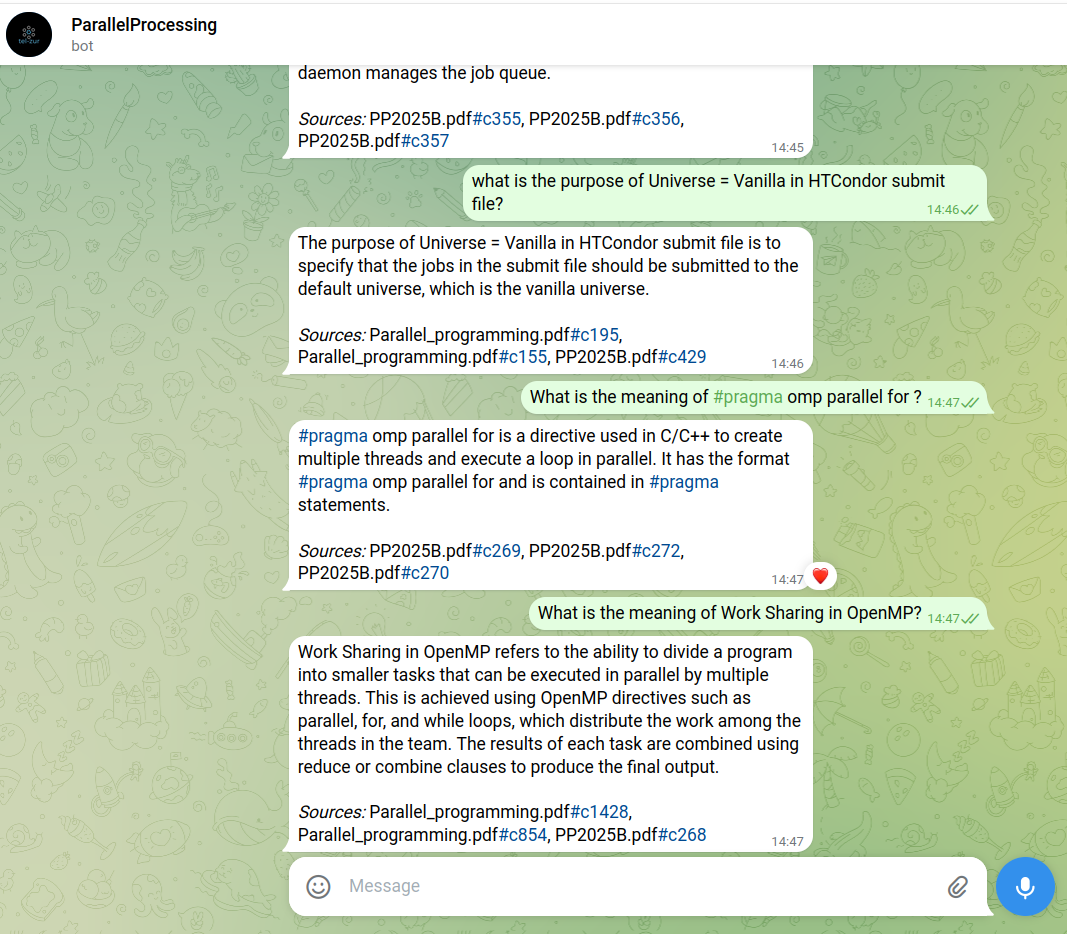}
    	\caption{The Telegram chat window.}
    	\label{fig:telegram-bot}
    \end{figure}

    A schematic chart showing all the project building blocks is shown in Figure \ref{fig:smart-agent-architecture}.
    A screen capture of the Telegram window showing a dialog between the user and the agent is shown in Figure \ref{fig:telegram-bot}.
    
    The Embeddings Generator converts each chunk of the course documents into numerical vector embeddings. This stage uses the open-source \textit{all-MiniLM-L6-v2} embedding model locally (via sentence-transformers),  ensuring privacy and low cost. The Vector Database (FAISS) stores the embeddings and their metadata for fast similarity search. It enables retrieval of the most relevant chunks based on semantic similarity to user queries. The Chunking and metadata indexing splits documents into manageable pieces (e.g. 512–1024 tokens) with overlap and keeps source references for later citation.
    The Retrieval-Augmented Generation (RAG) pipeline retrieves relevant information from the knowledge base and feed it to the LLM.
    The retriever accepts a user query and searches the vector DB for top-k most relevant chunks. Then it returns both the original text and source metadata.
    The context builder combines the retrieved chunks into a context prompt. It
    ensures the LLM receives only the most relevant and concise data.
    The local LLM inference (Mistral 7B) generates with the retrieved context coherent answers. Mistral 7B (Quantized GGUF) is loaded locally via llama-cpp-python with CUDA GPU offloading for fast inference.
    it uses both the user query and the retrieved context to craft accurate, grounded responses.
    The Telegram Bot provides an easy, conversational user interface.
    The bot back-end receives user messages from Telegram and forwards them to the RAG + LLM pipeline. It is also responsible to the response handling, i.e. sending back the generated answer.
    Finally, there is the orchestration and containerization which make the deployment reproducible and portable. I used \textit{Docker} and \textit{docker-compose}. This framework encapsulates all dependencies (Python, CUDA libraries, llama-cpp-python, FAISS). Using one-command there is a startup of the entire stack. In addition there is an environment configuration .env file that controls model path, GPU layer settings, and runtime parameters.
    
    The following paragraph adds some more details about the project building blocks:
        
    The core components include:
    \newline $\bullet$A Sentence-Transformers ('all-MiniLM-L6-v2')\cite{website:minilm} for embedding course content.
    \newline $\bullet$A FAISS vector store for fast semantic retrieval.
    \newline $\bullet$A quantized GGUF-format Mistral-7B model accelerated by llama.cpp and CUDA\cite{website:mistral}requiring  $\sim 4.07$ GiB of disk space.
    \newline $\bullet$A Telegram bot interface to support real-time interaction.
    
    all-MiniLM-L6-v2 is a specialized, compact, and efficient AI model that takes a sentence or short paragraph and converts it into a string of numbers, called a \emph{vector embedding}, which captures its semantic meaning. This process allows computers to understand and compare the meaning of different pieces of text, not just the words themselves.
    
    The model takes a sentence like "I love dogs" and another like "I adore canines." Even though the words are different, the model understands they have a similar meaning. It will then generate two numerical vectors that are very close to each other. On the other hand, the vector for "I love food" would be much farther away. 
    
    This ability to turn meaning into numbers makes all-MiniLM-L6-v2 useful for several tasks:
    
    Semantic Search: Instead of just finding documents with keywords, one can search for a concept. For example, a search for "places to get coffee" could return results for "cafes in my area."
    Clustering: It can automatically group similar pieces of text together. One could give it a list of customer reviews and it would sort them into groups like "reviews about customer service" and "reviews about product quality."
    Sentence Similarity: One can use it to find how similar two sentences are to each other, which is helpful for things like finding duplicate questions in a forum.
    
    The name "MiniLM" highlights its key advantage: it's a much smaller and faster version of larger, more powerful models. This makes it ideal for applications that need to run quickly or on devices with limited resources, like a mobile app or an embedded system. The "L6" means it has 6 transformer layers, and "v2" indicates it's an updated, improved version.
    
    A FAISS (Facebook AI Similarity Search) vector store is a database designed to store and quickly search through large collections of numerical representations of data, known as vectors. Think of it as a highly specialized, incredibly fast search engine for abstract concepts rather than keywords.
    
    \cite{website:mistral} and the associated paper \citealp{mistral-paper-2023} describe Mistral 7B, a large language model (LLM) which is developed by the French company Mistral AI. The "7B" in its name indicates that the model has 7.3 billion parameters. It's known for being powerful and efficient for its size, making it a strong competitor to larger models for certain tasks. Mistral 7B is also notable for its permissive license, which allows for free use and modifications. 
    After the RAG system retrieves the most relevant chunks from the knowledge base, it sends this retrieved context together with the user’s query to the language model. Mistral 7B then uses both the query and the retrieved context to generate a complete, coherent, and contextually grounded response. Because of its relatively small size compared to larger models, Mistral 7B can run efficiently on local hardware while still producing high-quality answers.
    The role of Mistral 7B is not performing the retrieval itself — that’s done by the RAG pipeline — but it integrates the retrieved data into its answer generation. Regarding its efficiency: Calling out its small size explains why it is practical to use locally and gives good response times on GPU.
    The output quality: since Mistral 7B is a general-purpose LLM, so it can combine the retrieved knowledge with its own contextual reasoning ability to produce complete answers.
    
    Telegram bot is an automated program that runs inside the Telegram messaging app. It's an account controlled by software, not a person, and it can perform various tasks like answering questions as in this project. The Telegram bot is created by "Bot Father" which is a special Telegram bot that is used to create and manage other bots like the one described in this project. When working with Bot Father one should first create the bot using the \textit{/newbot} command. Then a unique access token is generated (which must be kept secretly) is created after invoking the \textit{/token} command. This token must be available to the container as an  environment variable to be defined inside the .env file. Finally, there is a \textit{/setname} command to set the bot name.
     
    In order to successfully build the project a very delicate dependency structure between the various packages must be kept. In order to do that and to preserve reproducibility a \textit{requirement.txt} file is a must.
    Building the container is done as follows:
    
    \texttt{docker compose build --no-cache --build-arg CUDA\_ARCH=89}
    
    \noindent and then running it from the project root directory:
    
    \texttt{docker compose up -d}
    
\begin{figure}[htbp]
    	\centering
    	\includegraphics[scale=0.40]{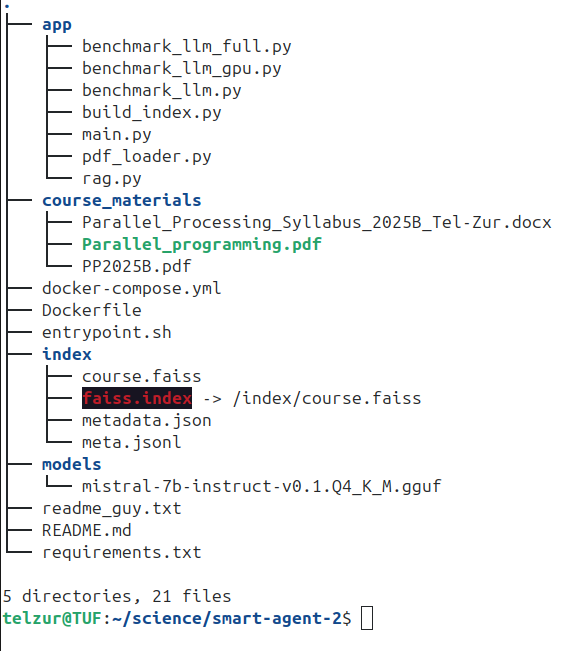}
    	\caption{The project tree structure.}
    	\label{fig:tree}
    \end{figure}

    The project tree is simple and straight forward. It is built in a standard structure suitable for using docker, see Figure \ref{fig:tree}.
    
    In order to verify that the container is running while using the GPU and is connected to telegram one could check the log file - see Appendix \ref*{app:logfile}.
    {\scriptsize \begin{verbatim}
    
    \end{verbatim}
}

\begin{figure}[htbp]
    	\centering
    	\includegraphics[scale=0.13]{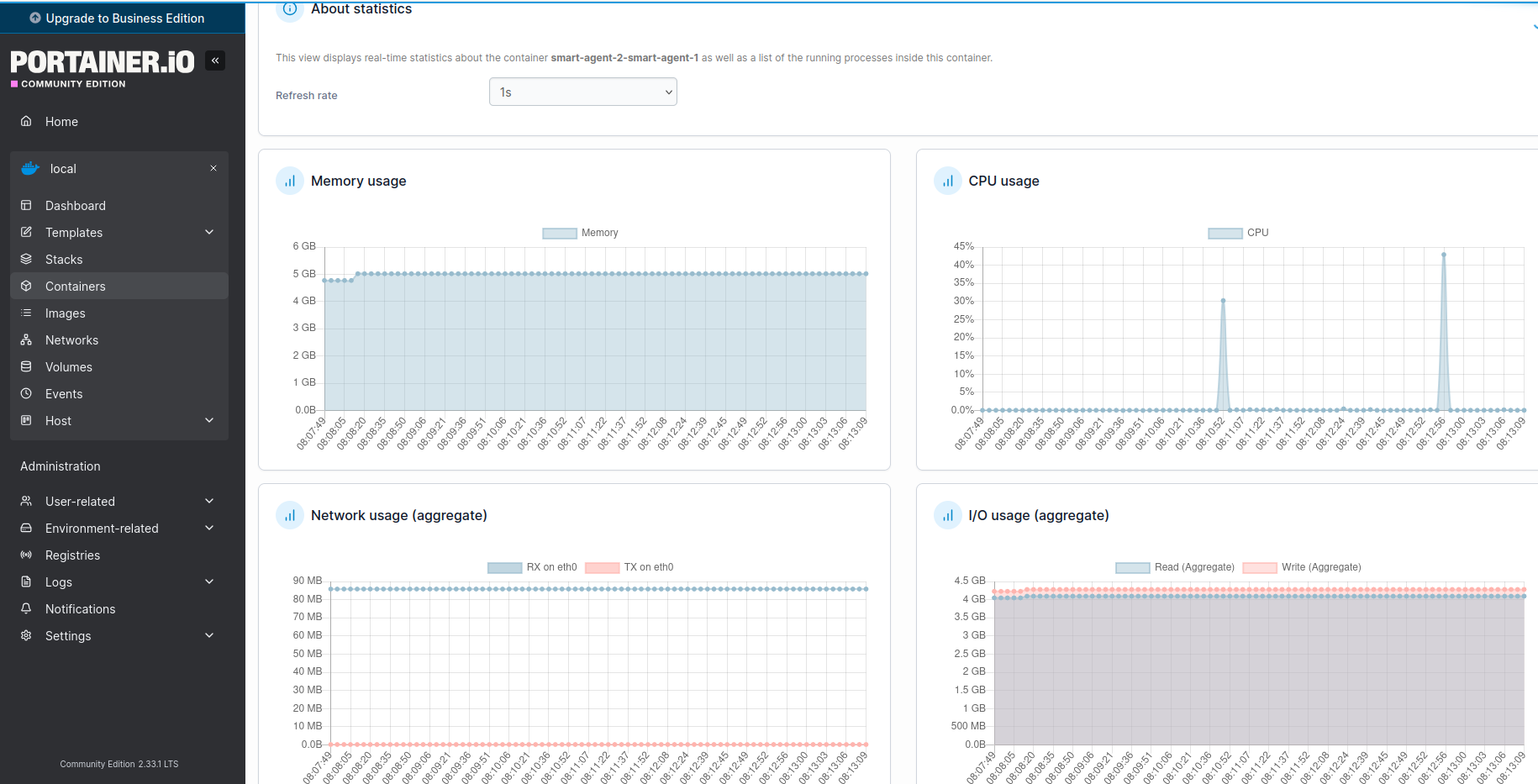}
    	\caption{The Portainer containers management system.}
    	\label{fig:portainer}
    \end{figure}

    It is recommended but not mandatory to use in this and alike projects a container manager such as Portainer\cite{website:portainer} which allows better control on the containers management and control. Figure \ref{fig:portainer} shows the container load. The two peaks seen in the CPU load graph correspond to two queries.
    
    \paragraph{Benchmarking}
    \begin{quotation}
    \textit{"What you can't measure doesn't exist"}
    \end{quotation}
    The most relevant performance factor for this system is the number of Tokens Per Second (TPS).
		
	Let us compare the estimated performance on 3 platforms 1) A typical laptop running the model on the CPU 2) A laptop running on the GPU, and 3) A cloud server such as AWS g5.xlarge with a stronger GPU. Where all of them use the Mistral 7B model.
	
	{\scriptsize 
	\begin{tabular}{|p{0.7cm}|p{3.5cm}|l|}
		\hline
		\textbf{System} & \textbf{Description} & \textbf{Estimated TPS} \\
		\hline
		1 & A typical laptop running on the CPU & $\sim 0.5-1.5$ \\
		\hline
		2 & This system, a laptop running on the GPU  (Nividia GeForce RTX  4060 with 8GiB) & about 16 (measured in a benchmark) \\
		\hline
		3 & A cloud server such as AWS like g5.xlarge which has 4 vCPUs, 16GiB memory and Nvidia A10 architecture with 24GiB memory but with an average on-demand cost of \$1.22 per hour! & ~30-100 \\
		\hline
	\end{tabular}
    }
    
    In Appendix \ref*{app:logfile} enclosed a log file with information about the container performance. Let us now give an interpretation of the results on the RTX 4060 Laptop GPU (8 GB VRAM) and Mistral-7B Q4\_K\_M model.
    
    The mean generation speed, gen\_tps $\sim 16$tokens/s. The total TPS is close to the generation TPS.
    Another important factor is the time from request to first token (first token latency) which is called for short as TTFB, The mean TTFB is about $\sim 0.1$s.
    The total latency can be estimated according to
    $$ T_{total}\approx T_{TTFB}+\frac{N_{out}}{R_{gen}} $$
    where $N_{out}$ is the number of tokens generated and $R_{gen}$ is the generation speed (toekns per second, gen\_tps in the code).
    For example, if we assume $T_{TTFB}=0.1$s, $R_{gen}=16 $ toekens/s, and $N_{out}-150 $ tokens, we get $T_{total}\approx 9.5$s.    
    
    Summary
    The RTX 4060 Laptop GPU is performing as expected: $\approx0.1$s TTFB and $\approx 16$ tokens/sec on a quantized 7B model. That’s good for smooth chatbots, teaching demos, or personal assistants. To scale to more users or bigger contexts, there will be a  need for more VRAM (desktop 4090/4080, or cloud A100).  
    
    The smart agent performance benchmark shows that it feels responsive and can serve as a production-ready on a RTX 4060 laptop GPU!
	
	\section{Portability and Reproducibility}
	All the codes and instructions how to build the project are available at: \url{https://tel-zur.net/papers/EduHPC25}. 
    In order to save space Mistral 7B is absent from this package and has to be separately downloaded from the Mistral website\cite{website:mistral}. 
	In order to allow the container to use the GPU we must first install on the host computer the Nvidia container toolkit which can be done by following the next 3 steps:
	
	{\scriptsize \begin{verbatim}
		# 1. Add the NVIDIA GPG key
		curl -fsSL https://nvidia.github.io/libnvidia-container/gpgkey | \
		sudo gpg --dearmor -o /usr/share/keyrings/nvidia-container-toolkit.gpg
		
		# 2. Add the correct repo (for Ubuntu 24.04 -> "ubuntu24.04")
		distribution=\$(. /etc/os-release; echo \${ID}\${VERSION_ID})
		curl -s -L https://nvidia.github.io/libnvidia-container/ubuntu24.04/\
		libnvidia-container.list | \\
		sed 's#deb https://#deb [signed-by=/usr/share/keyrings/nvidia-container-toolkit.gpg]\
		 https://#g' | sudo tee /etc/apt/sources.list.d/nvidia-container-toolkit.list
		
		# 3. Update and install
		sudo apt update
		sudo apt install -y nvidia-container-toolkit
	\end{verbatim} 
}
	
	\section{Future work}
	
	The current system performs quite good but it can be further optimized by fine-tuning several parameters.
	
	1. Batch tuning: Tuning the number of batches. Larger n\_batch can improve throughput but may hit VRAM limits.
	Currently I used n\_batch=256. 
	
	2. The  number of GPU layers. The number of transformer layers (out of the total model depth) to offload from CPU to the GPU. A 7B model like Mistral has 32 transformer layers.
	For example if this number is set to be 32, it tries to put all layers on GPU. However, there is a trade-off, on one had more layers on GPU has a faster inference, on the other hand more layers means more VRAM usage.
	In this test using the RTX 4060 GPU full 32 layers may exceeds VRAM. The sweet spot should is around 20 layers on GPU, and the rest on CPU. The value can be seen in the log file in the appendix.
	$N\_GPU\_LAYERS =0$ means that we use only the CPU (lowest VRAM and slowest), $10-20$ is a mixed CPU-GPU and is a good compromise, $=32+$ means a full GPU offload. This mode needs $\sim 4-5 $ GiB just for the weights on a Q4\_K 7B model and additional VRAM for the KV and the compute, which might cause Out of Memeory (OOM).
		
	3. Tensor split: Fine-grained control over how model tensors are divided across multiple GPUs. 
	In this project I use only a single GPU and the tensor split value is 0.85, which balances between the CPU (0.15) and the GPU (0.85).  For a single GPU setups, tensor split values are largely irrelevant, but they become important in multi-GPU systems. Tensor split value is configured in \textit{llama.cpp} and \textit{lamma-cpp-python}.
	
	4. The maximum context window \textit{n\_ctx} in the code which is the number of tokens the model can "see" at once. 
	\textit{n\_ctx} controls the size of the Key-Value (KV)-cache that stores intermediate key/value tensors for the attention mechanism.
	If, for example, a token is usually 3-4 characters of text it means that 1024 tokens are equivalent to about 700-800 words. In the present code we set \textit{n\_ctx=768} which means the model can consider up to 768 tokens (prompt and generated output together) in one pass which is roughly about 600 words. This parameter should also be tuned, the higher \textit{n\_ctx} the higher the memory usage will be (VRAM and CPU RAM) and the hight the smart agent can reason over long context. On the other hand small \textit{n\_ctx} results in a shorter VRAM and it might lose context on long prompt but it will react faster because it will compute less.
	As a rule of thumb, increase \textit{n\_ctx} only when you see the model truncating context (answers ignoring retrieved text), otherwise, keep it small to stay fast.
	
	5. Low VRAM. Tuning all the above parameter should be done such that we will always avoid Out of Memeory (OOM) when running long contexts.
	
	6. Driver + build improvements:
	Newer llama.cpp builds (compiled with CUDA graph optimizations).
	Setting \textit{flash\_attn=True} in \textit{llama\_cpp} tells it to use Flash Attention kernels instead of the standard attention implementation. 	
	Flash Attention is a highly optimized CUDA kernel that computes attention in $O(n^2)$ time but with $O(n)$ memory, by streaming the computation and avoiding storing the full attention matrix.	
	This gives a big speedup and lower peak VRAM usage on supported GPUs, especially for long context sizes and larger batch sizes.
	flash\_attention is supported on compute capability $\geq 7.0$ (Volta and newer) The RTX4060 has a compute capability of 8.9 and supports it.
	
	7. Quantization of the model. This affects the number of bits per weight. The lower this number the less VRAM it takes. Currently we use Q4\_K\_M which seems to be a good balance. Q3\_K\_M may reduce VRAM further but at the cost of quality degradation.
	
    To sum up the current setting which is throughput focused is:
    N\_GPU\_LAYERS=20, N\_CTX=768, N\_BATCH=256, FLASH\_ATTN=1. 
    These optimizations form a roadmap for scaling the system to support more users, longer contexts, and larger modules.
    
	\section{Conclusions} 
    The smart agent for the "Introduction to Parallel Processing" course is ready. Initial implementation will take place in the forthcoming semester. Probably this will be followed by an improvement cycle. A departmental server is the right place to host the agent and I hope that such a server will be allocated to the project. Additional courses can easily be added. At this point of time the right way to conclude this paper is to cite Winston Churchill who said: 'Now this is not the end. It is not even the beginning of the end. But it is, perhaps, the end of the beginning.'
    
	\bibliographystyle{ACM-Reference-Format}
	\bibliography{Tel-Zur_bib}
	
	\appendix
	\section{Glossary}
	$\bullet$PTB - Python-Telegram-Bot (version 20.7 in this project)
	\newline $\bullet$TTFB - Time To First Byte, how long it takes from sending the user prompt until the model starts producing the very first output token.
	\newline $\bullet$Prompt length - the number of characters in the raw text sent (input).
	\newline $\bullet$Prompt tokens - the number of sub-word units used internally by the model (the number of token is relevant when asking about the VRAM usage)
	\newline $\bullet$Completion tokens - the number of tokens the model generated as output.
	\newline $\bullet$Generation duration - time spent generating the output tokens after the first one arrived (i.e. excluding TTFB).
	\newline $\bullet$Total duration - The entire end to end time from sending the prompt until the last token of the output, i.e. Total duration = TTFB + Generation duration.
	\newline $\bullet$TPS - Tokens per second is the  throughput.
	Generation TPS = completion tokens + Generation duration (a measurement of the Chat-bot responsiveness)
	\newline $\bullet$Total TPS = prompt tokens + completion tokens + tutal duration. This is the whole transaction throughput: how much useful work is done over the entire request. 
	Max. Token = The upper limit to the output length, e.g. 128, which means that the model will not exceed 128 tokens in the output.
	\newline $\bullet$Temperature - The randomness setting (low 0.1-0.3, medium 0.7-1.0, high <1)
		
	\onecolumn
	\section{Logfile}
	\label{app:logfile}
	I enclose here a logfile of the the running container where important information about the CPU-GPU offloading can be seen together with some performance results:
	{\tiny\begin{verbatim}
telzur@TUF:~/science/smart-agent-2$ docker compose run --rm -T \
-e N_GPU_LAYERS=20 \
-e N_BATCH=256 \
-e N_CTX=768 \
--entrypoint python smart-agent /app/benchmark_llm_full.py
== LLM config ==
model_path=/models/mistral-7b-instruct-v0.1.Q4_K_M.gguf
n_ctx=768  n_batch=256  n_gpu_layers=20  flash_attn=True
ggml_cuda_init: GGML_CUDA_FORCE_MMQ:    yes
ggml_cuda_init: GGML_CUDA_FORCE_CUBLAS: no
ggml_cuda_init: found 1 CUDA devices:
Device 0: NVIDIA GeForce RTX 4060 Laptop GPU, compute capability 8.9, VMM: yes
llama_model_load_from_file_impl: using device CUDA0 (NVIDIA GeForce RTX 4060 Laptop GPU) - 7622 MiB free
llama_model_loader: loaded meta data with 20 key-value pairs and 291 tensors from /models/mistral-7b-instruct-v0.1.Q4_K_M.gguf (version GGUF V2)
llama_model_loader: Dumping metadata keys/values. Note: KV overrides do not apply in this output.
llama_model_loader: - kv   0:                       general.architecture str              = llama
llama_model_loader: - kv   1:                               general.name str              = mistralai_mistral-7b-instruct-v0.1
llama_model_loader: - kv   2:                       llama.context_length u32              = 32768
llama_model_loader: - kv   3:                     llama.embedding_length u32              = 4096
llama_model_loader: - kv   4:                          llama.block_count u32              = 32
llama_model_loader: - kv   5:                  llama.feed_forward_length u32              = 14336
llama_model_loader: - kv   6:                 llama.rope.dimension_count u32              = 128
llama_model_loader: - kv   7:                 llama.attention.head_count u32              = 32
llama_model_loader: - kv   8:              llama.attention.head_count_kv u32              = 8
llama_model_loader: - kv   9:     llama.attention.layer_norm_rms_epsilon f32              = 0.000010
llama_model_loader: - kv  10:                       llama.rope.freq_base f32              = 10000.000000
llama_model_loader: - kv  11:                          general.file_type u32              = 15
llama_model_loader: - kv  12:                       tokenizer.ggml.model str              = llama
llama_model_loader: - kv  13:                      tokenizer.ggml.tokens arr[str,32000]   = ["<unk>", "<s>", "</s>", "<0x00>", "<...
llama_model_loader: - kv  14:                      tokenizer.ggml.scores arr[f32,32000]   = [0.000000, 0.000000, 0.000000, 0.0000...
llama_model_loader: - kv  15:                  tokenizer.ggml.token_type arr[i32,32000]   = [2, 3, 3, 6, 6, 6, 6, 6, 6, 6, 6, 6, ...
llama_model_loader: - kv  16:                tokenizer.ggml.bos_token_id u32              = 1
llama_model_loader: - kv  17:                tokenizer.ggml.eos_token_id u32              = 2
llama_model_loader: - kv  18:            tokenizer.ggml.unknown_token_id u32              = 0
llama_model_loader: - kv  19:               general.quantization_version u32              = 2
llama_model_loader: - type  f32:   65 tensors
llama_model_loader: - type q4_K:  193 tensors
llama_model_loader: - type q6_K:   33 tensors
print_info: file format = GGUF V2
print_info: file type   = Q4_K - Medium
print_info: file size   = 4.07 GiB (4.83 BPW) 
init_tokenizer: initializing tokenizer for type 1
load: control token:      2 '</s>' is not marked as EOG
load: control token:      1 '<s>' is not marked as EOG
load: special_eos_id is not in special_eog_ids - the tokenizer config may be incorrect
load: printing all EOG tokens:
load:   - 2 ('</s>')
load: special tokens cache size = 3
load: token to piece cache size = 0.1637 MB
print_info: arch             = llama
print_info: vocab_only       = 0
print_info: n_ctx_train      = 32768
print_info: n_embd           = 4096
print_info: n_layer          = 32
print_info: n_head           = 32
print_info: n_head_kv        = 8
print_info: n_rot            = 128
print_info: n_swa            = 0
print_info: is_swa_any       = 0
print_info: n_embd_head_k    = 128
print_info: n_embd_head_v    = 128
print_info: n_gqa            = 4
print_info: n_embd_k_gqa     = 1024
print_info: n_embd_v_gqa     = 1024
print_info: f_norm_eps       = 0.0e+00
print_info: f_norm_rms_eps   = 1.0e-05
print_info: f_clamp_kqv      = 0.0e+00
print_info: f_max_alibi_bias = 0.0e+00
print_info: f_logit_scale    = 0.0e+00
print_info: f_attn_scale     = 0.0e+00
print_info: n_ff             = 14336
print_info: n_expert         = 0
print_info: n_expert_used    = 0
print_info: causal attn      = 1
print_info: pooling type     = 0
print_info: rope type        = 0
print_info: rope scaling     = linear
print_info: freq_base_train  = 10000.0
print_info: freq_scale_train = 1
print_info: n_ctx_orig_yarn  = 32768
print_info: rope_finetuned   = unknown
print_info: model type       = 7B
print_info: model params     = 7.24 B
print_info: general.name     = mistralai_mistral-7b-instruct-v0.1
print_info: vocab type       = SPM
print_info: n_vocab          = 32000
print_info: n_merges         = 0
print_info: BOS token        = 1 '<s>'
print_info: EOS token        = 2 '</s>'
print_info: UNK token        = 0 '<unk>'
print_info: LF token         = 13 '<0x0A>'
print_info: EOG token        = 2 '</s>'
print_info: max token length = 48
load_tensors: loading model tensors, this can take a while... (mmap = true)
load_tensors: layer   0 assigned to device CPU, is_swa = 0
load_tensors: layer   1 assigned to device CPU, is_swa = 0
load_tensors: layer   2 assigned to device CPU, is_swa = 0
load_tensors: layer   3 assigned to device CPU, is_swa = 0
load_tensors: layer   4 assigned to device CPU, is_swa = 0
load_tensors: layer   5 assigned to device CPU, is_swa = 0
load_tensors: layer   6 assigned to device CPU, is_swa = 0
load_tensors: layer   7 assigned to device CPU, is_swa = 0
load_tensors: layer   8 assigned to device CPU, is_swa = 0
load_tensors: layer   9 assigned to device CPU, is_swa = 0
load_tensors: layer  10 assigned to device CPU, is_swa = 0
load_tensors: layer  11 assigned to device CPU, is_swa = 0
load_tensors: layer  12 assigned to device CUDA0, is_swa = 0
load_tensors: layer  13 assigned to device CUDA0, is_swa = 0
load_tensors: layer  14 assigned to device CUDA0, is_swa = 0
load_tensors: layer  15 assigned to device CUDA0, is_swa = 0
load_tensors: layer  16 assigned to device CUDA0, is_swa = 0
load_tensors: layer  17 assigned to device CUDA0, is_swa = 0
load_tensors: layer  18 assigned to device CUDA0, is_swa = 0
load_tensors: layer  19 assigned to device CUDA0, is_swa = 0
load_tensors: layer  20 assigned to device CUDA0, is_swa = 0
load_tensors: layer  21 assigned to device CUDA0, is_swa = 0
load_tensors: layer  22 assigned to device CUDA0, is_swa = 0
load_tensors: layer  23 assigned to device CUDA0, is_swa = 0
load_tensors: layer  24 assigned to device CUDA0, is_swa = 0
load_tensors: layer  25 assigned to device CUDA0, is_swa = 0
load_tensors: layer  26 assigned to device CUDA0, is_swa = 0
load_tensors: layer  27 assigned to device CUDA0, is_swa = 0
load_tensors: layer  28 assigned to device CUDA0, is_swa = 0
load_tensors: layer  29 assigned to device CUDA0, is_swa = 0
load_tensors: layer  30 assigned to device CUDA0, is_swa = 0
load_tensors: layer  31 assigned to device CUDA0, is_swa = 0
load_tensors: layer  32 assigned to device CPU, is_swa = 0
load_tensors: tensor 'token_embd.weight' (q4_K) (and 110 others) cannot be used with preferred buffer type CUDA_Host, using CPU instead
load_tensors: offloading 20 repeating layers to GPU
load_tensors: offloaded 20/33 layers to GPU
load_tensors:        CUDA0 model buffer size =  2495.31 MiB
load_tensors:   CPU_Mapped model buffer size =  4165.37 MiB
...............................................................................................
llama_context: constructing llama_context
llama_context: n_seq_max     = 1
llama_context: n_ctx         = 768
llama_context: n_ctx_per_seq = 768
llama_context: n_batch       = 256
llama_context: n_ubatch      = 256
llama_context: causal_attn   = 1
llama_context: flash_attn    = 1
llama_context: kv_unified    = false
llama_context: freq_base     = 10000.0
llama_context: freq_scale    = 1
llama_context: n_ctx_per_seq (768) < n_ctx_train (32768) -- the full capacity of the model will not be utilized
set_abort_callback: call
llama_context:        CPU  output buffer size =     0.12 MiB
create_memory: n_ctx = 768 (padded)
llama_kv_cache_unified: layer   0: dev = CPU
llama_kv_cache_unified: layer   1: dev = CPU
llama_kv_cache_unified: layer   2: dev = CPU
llama_kv_cache_unified: layer   3: dev = CPU
llama_kv_cache_unified: layer   4: dev = CPU
llama_kv_cache_unified: layer   5: dev = CPU
llama_kv_cache_unified: layer   6: dev = CPU
llama_kv_cache_unified: layer   7: dev = CPU
llama_kv_cache_unified: layer   8: dev = CPU
llama_kv_cache_unified: layer   9: dev = CPU
llama_kv_cache_unified: layer  10: dev = CPU
llama_kv_cache_unified: layer  11: dev = CPU
llama_kv_cache_unified: layer  12: dev = CUDA0
llama_kv_cache_unified: layer  13: dev = CUDA0
llama_kv_cache_unified: layer  14: dev = CUDA0
llama_kv_cache_unified: layer  15: dev = CUDA0
llama_kv_cache_unified: layer  16: dev = CUDA0
llama_kv_cache_unified: layer  17: dev = CUDA0
llama_kv_cache_unified: layer  18: dev = CUDA0
llama_kv_cache_unified: layer  19: dev = CUDA0
llama_kv_cache_unified: layer  20: dev = CUDA0
llama_kv_cache_unified: layer  21: dev = CUDA0
llama_kv_cache_unified: layer  22: dev = CUDA0
llama_kv_cache_unified: layer  23: dev = CUDA0
llama_kv_cache_unified: layer  24: dev = CUDA0
llama_kv_cache_unified: layer  25: dev = CUDA0
llama_kv_cache_unified: layer  26: dev = CUDA0
llama_kv_cache_unified: layer  27: dev = CUDA0
llama_kv_cache_unified: layer  28: dev = CUDA0
llama_kv_cache_unified: layer  29: dev = CUDA0
llama_kv_cache_unified: layer  30: dev = CUDA0
llama_kv_cache_unified: layer  31: dev = CUDA0
llama_kv_cache_unified:      CUDA0 KV buffer size =    60.00 MiB
llama_kv_cache_unified:        CPU KV buffer size =    36.00 MiB
llama_kv_cache_unified: size =   96.00 MiB (   768 cells,  32 layers,  1/1 seqs), K (f16):   48.00 MiB, V (f16):   48.00 MiB
llama_context: enumerating backends
llama_context: backend_ptrs.size() = 2
llama_context: max_nodes = 2328
llama_context: worst-case: n_tokens = 256, n_seqs = 1, n_outputs = 0
graph_reserve: reserving a graph for ubatch with n_tokens =  256, n_seqs =  1, n_outputs =  256
graph_reserve: reserving a graph for ubatch with n_tokens =    1, n_seqs =  1, n_outputs =    1
graph_reserve: reserving a graph for ubatch with n_tokens =  256, n_seqs =  1, n_outputs =  256
llama_context:      CUDA0 compute buffer size =   137.79 MiB
llama_context:  CUDA_Host compute buffer size =     4.76 MiB
llama_context: graph nodes  = 999
llama_context: graph splits = 136 (with bs=256), 3 (with bs=1)
CUDA : ARCHS = 500,520,530,600,610,620,700,720,750,800,860,870,890,900 | FORCE_MMQ = 1 | USE_GRAPHS = 1 | PEER_MAX_BATCH_SIZE = 128 | CPU : SSE3 = 1 | SSSE3 = 1 | AVX = 1 | AVX2 = 1 | F16C = 1 | FMA = 1 | BMI2 = 1 | LLAMAFILE = 1 | OPENMP = 1 | REPACK = 1 | 
Model metadata: {'tokenizer.ggml.unknown_token_id': '0', 'tokenizer.ggml.eos_token_id': '2', 'general.architecture': 'llama', 'llama.rope.freq_base': '10000.000000', 'llama.context_length': '32768', 'general.name': 'mistralai_mistral-7b-instruct-v0.1', 'llama.embedding_length': '4096', 'llama.feed_forward_length': '14336', 'llama.attention.layer_norm_rms_epsilon': '0.000010', 'llama.rope.dimension_count': '128', 'tokenizer.ggml.bos_token_id': '1', 'llama.attention.head_count': '32', 'llama.block_count': '32', 'llama.attention.head_count_kv': '8', 'general.quantization_version': '2', 'tokenizer.ggml.model': 'llama', 'general.file_type': '15'}
Using fallback chat format: llama-2
llama_perf_context_print:        load time =     530.16 ms
llama_perf_context_print: prompt eval time =     530.09 ms /     3 tokens (  176.70 ms per token,     5.66 tokens per second)
llama_perf_context_print:        eval time =     380.94 ms /     7 runs   (   54.42 ms per token,    18.38 tokens per second)
llama_perf_context_print:       total time =     913.28 ms /    10 tokens
llama_perf_context_print:    graphs reused =          6

== Running ==
Llama.generate: 1 prefix-match hit, remaining 23 prompt tokens to eval
llama_perf_context_print:        load time =     530.16 ms
llama_perf_context_print: prompt eval time =     349.16 ms /    23 tokens (   15.18 ms per token,    65.87 tokens per second)
llama_perf_context_print:        eval time =    6852.83 ms /   127 runs   (   53.96 ms per token,    18.53 tokens per second)
llama_perf_context_print:       total time =    7237.84 ms /   150 tokens
llama_perf_context_print:    graphs reused =        126
[1/5] TTFB=0.350s | gen_tps=16.99 | total=7.24s | comp_tok~117 | GPU sm% avg/max=21.7/23 | GPU mem MB avg/max=2903.3/2904
Llama.generate: 23 prefix-match hit, remaining 1 prompt tokens to eval
llama_perf_context_print:        load time =     530.16 ms
llama_perf_context_print: prompt eval time =       0.00 ms /     1 tokens (    0.00 ms per token,      inf tokens per second)
llama_perf_context_print:        eval time =    6857.01 ms /   128 runs   (   53.57 ms per token,    18.67 tokens per second)
llama_perf_context_print:       total time =    6891.17 ms /   129 tokens
llama_perf_context_print:    graphs reused =        128
[2/5] TTFB=0.067s | gen_tps=16.27 | total=6.89s | comp_tok~111 | GPU sm% avg/max=21.8/23 | GPU mem MB avg/max=2904.0/2904
Llama.generate: 23 prefix-match hit, remaining 1 prompt tokens to eval
llama_perf_context_print:        load time =     530.16 ms
llama_perf_context_print: prompt eval time =       0.00 ms /     1 tokens (    0.00 ms per token,      inf tokens per second)
llama_perf_context_print:        eval time =    6881.64 ms /   128 runs   (   53.76 ms per token,    18.60 tokens per second)
llama_perf_context_print:       total time =    6915.33 ms /   129 tokens
llama_perf_context_print:    graphs reused =        128
[3/5] TTFB=0.062s | gen_tps=15.61 | total=6.92s | comp_tok~107 | GPU sm% avg/max=21.5/23 | GPU mem MB avg/max=2904.0/2904
Llama.generate: 23 prefix-match hit, remaining 1 prompt tokens to eval
llama_perf_context_print:        load time =     530.16 ms
llama_perf_context_print: prompt eval time =       0.00 ms /     1 tokens (    0.00 ms per token,      inf tokens per second)
llama_perf_context_print:        eval time =    6726.33 ms /   126 runs   (   53.38 ms per token,    18.73 tokens per second)
llama_perf_context_print:       total time =    6760.74 ms /   127 tokens
llama_perf_context_print:    graphs reused =        126
[4/5] TTFB=0.062s | gen_tps=16.12 | total=6.76s | comp_tok~108 | GPU sm% avg/max=21.5/23 | GPU mem MB avg/max=2904.0/2904
Llama.generate: 23 prefix-match hit, remaining 1 prompt tokens to eval
llama_perf_context_print:        load time =     530.16 ms
llama_perf_context_print: prompt eval time =       0.00 ms /     1 tokens (    0.00 ms per token,      inf tokens per second)
llama_perf_context_print:        eval time =    6821.11 ms /   128 runs   (   53.29 ms per token,    18.77 tokens per second)
llama_perf_context_print:       total time =    6856.37 ms /   129 tokens
llama_perf_context_print:    graphs reused =        128
[5/5] TTFB=0.065s | gen_tps=16.05 | total=6.86s | comp_tok~109 | GPU sm% avg/max=21.1/23 | GPU mem MB avg/max=2904.0/2904

=== SUMMARY ===
{
	"model": "/models/mistral-7b-instruct-v0.1.Q4_K_M.gguf",
	"n_ctx": 768,
	"n_batch": 256,
	"n_gpu_layers": 20,
	"flash_attn": true,
	"prompt": "In one paragraph, explain what GPU offloading does in llama.cpp and why it speed\u2026",
	"iterations": 5,
	"metrics": {
		"ttfb_s": {
			"mean": 0.121,
			"median": 0.065,
			"p95": 0.067,
			"min": 0.062,
			"max": 0.35
		},
		"gen_tps": {
			"mean": 16.21,
			"median": 16.12,
			"p95": 16.265,
			"min": 15.61,
			"max": 16.99
		},
		"total_latency_s": {
			"mean": 6.933,
			"median": 6.892,
			"p95": 6.916,
			"min": 6.761,
			"max": 7.238
		},
		"gpu_util_sm_pct": {
			"mean": 21.5,
			"max": 23
		},
		"gpu_mem_mb": {
			"mean": 2903.9,
			"max": 2904
		}
	}
}
	\end{verbatim}
}
\end{document}